\documentclass[twocolumn, prb, showpacs]{revtex4}

\usepackage{amssymb}
\usepackage{graphicx}
\usepackage{dcolumn}
\usepackage{bm}
\usepackage{amsmath}
\begin{document}
\title{D-wave bosonic pair in an optical lattice}

\author{Zi Cai$^1$, Lei Wang$^2$ , Jian Li$^3$, Shu Chen$^2$ X. C. Xie$^{4,2,5}$,
, Yupeng Wang$^2$}

\affiliation{$^{1}$Department of Physics, University of California,
San Diego, California 92093, USA}

\affiliation{$^{2}$ Beijing National Laboratory for Condensed Matter
Physics, Institute of Physics, Chinese Academy of Sciences, Beijing
100080, P. R. China}

\affiliation{$^{3}$ Texas Center for Superconductivity and
Department of Physics, University of Houston, Houston, Texas 77204,
USA}

\affiliation{$^{4}$International Center for Quantum Materials,
Peking University, Beijing 100871, China}

\affiliation{$^{5}$Department of Physics, Oklahoma State University,
Stillwater, Oklahoma 74078, USA}

\date{ \today }

\begin{abstract}
We present a bosonic model, in which two bosons may form a bound
pair with d-wave symmetry via the four-site ring exchange
interaction. A d-wave pairing superfluid as well as a d-wave density
wave (DDW) state, are proposed to be achievable in this system. By
the mean field approach, we find that at low densities, the d-wave
pairs may condensate, leading to a d-wave bosonic paired superfluid.
At half filling, a d-wave Mott insulator could be realized in a
superlattice structure. At some particular filling factors, there
exists a novel phase: d-wave density wave state, which preserves the
d-wave symmetry within plaquette while spontaneously breaks the
translational symmetry. The DDW state and its corresponding quantum
phase transition in a two-leg ladder are studied by the
time-evolving block decimation (TEBD) method. We show that this
exotic bosonic system can be realized in the BEC zone of cold Fermi
gases loaded in a two-dimensional (2D) spin-dependent optical
lattice.
\end{abstract}

\pacs{05.30.Jp, 03.75.Nt, 74.20.Mn, 73.43.Nq }

\maketitle

Recently, ultracold atoms in optical lattice have provided a perfect
platform for simulating quantum many-body model in condensed matter
physics. Because of the flexible tunability of parameters such as
the hopping amplitudes, interaction or even the dimensionality of
the system, ultracold atomic systems allow us to directly study some
fundamental Hamiltonian systems and their associated phase
transitions, such as boson Hubbard model and the superfluid to Mott
insulator phase transition \cite{Greiner}, or the recent realization
of the repulsive or attractive fermionic Hubbard model
\cite{Jordens}. In addition, some exotic phases emerged from the
low-dimensional strongly correlated systems, such as resonating
valence bond (RVB) state\cite{Trebst,Paredes,Peterson}, d-wave
superfluidity\cite{Hofstetter,Honerkamp}, deconfined Coulomb phase
\cite{Buchler,Tewari} as well as  the topological insulator with
fractional statistic and topological order \cite{Paredes,Han}can
also be investigated in the cold atom systems. Furthermore, the
uniqueness of cold atomic system also provides new playgrounds for
physicists, such as the strongly correlated model for higher spin
systems, higher orbital systems \cite{Wu2,Zhao,Wang} or for the
optical superlattice \cite{Paredes,Paredes2,Damski,Goodman}. A
two-dimensional (2D) optical superlattice may be constructed by
imposing two optical lattices with different periods to form an
array of plaquettes.  The hopping amplitude and interaction for
atoms between these plaquettes are much smaller than that within the
plaquette. One of the exotic phase emerges in the superlattice is
the d-wave Mott insulator \cite{Yao,Trebst,Peterson} and d-wave
superfluid \cite{Rey}. The d-wave Mott insulator is the insulator
state with local d-wave symmetry, i.e., if we rotate the site within
a plaquette by $\pi/2$, the wave function reverses its sign. When we
introduce holes into the d-wave Mott insulator, two holes tend to
bind together within the plaquette to form a Cooper pair with local
d-wave symmetry, and the propagation of the d-wave pairs between
different plaques leads to the d-wave superfluid.

The mechanism of pairing with d-wave symmetry has played an
important role in the high-Tc superconductor. Though without
rigorous proof, numerous evidences strongly support the existence of
d-wave superconductor (or superfluidity in cold atomic system) near
the half-filling.  The background of Neel state with
antiferromagnetic correlations plays a key role in this mechanism of
d-wave symmetry \cite{Scalapino}. In cold fermionic atom system,
however, the binding energy of a d-wave pair is much smaller than
the hopping amplitude, which makes it difficult to directly simulate
the d-wave mechanism of the high-Tc superconductor in cold atomic
systems. One solution is to trap the d-wave pair within a plaquette
of the optical superlattice. In this paper, we propose a novel
mechanism for realizing the d-wave pairing. Different from that in
high-Tc superconductor induced by antiferromagnetic correlation, the
d-wave paring here is induced by the four-site ring exchange
interaction.

\begin{figure}[htb]
\includegraphics[width=8.5cm,bb=14 331 542 684]
{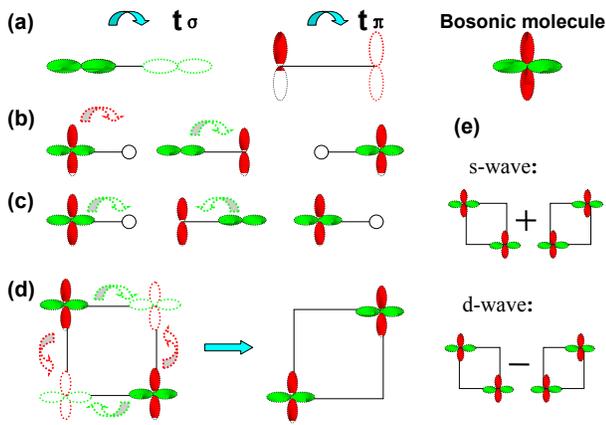} \caption{(a) Anisotropic hopping matrix elements in
the 2D anisotropic spin-dependent optical lattice (green $\uparrow$;
red $\downarrow$): $t_\pi$ and $t_\sigma$. (b) The effective hopping
term for the hardcore boson as a result of second order
perturbation. (c) The effective interaction term for the hardcore
boson. (d) The four-site ring exchange interaction for hardcore
boson as a result of fourth perturbation. (e) Two eigenstate s-wave
and d-wave symmetry within one plaquette} \label{fig1}
\end{figure}

Before discussing the physical realization of the effective
Hamiltonian, we first present the Hamiltonian, which is a hard-core
bosonic model with a strong nearest neighbor (NN) repulsive
interaction and a four-site ring interaction:
\begin{equation}
H=\sum_{\langle ij\rangle}[ta^\dag_i a_j+Vn_i
n_j]-\mu\sum_in_i+K\sum_{\langle ijkl\rangle} a^\dag_ia_ja^\dag_ka_l
+h.c\label{EQ1}
\end{equation}

where $\langle ij\rangle$ denotes a pair of nearest-neighbor sites
and $\langle ijkl\rangle$ are sites on the corners of a plaquette.
In this paper ,we focus on the parameters region  $V\gg K \gg
|t|>0$. The four-site ring exchange interaction with positive
coefficient (K$>$0) may be very important in determining the
properties of this system (at least at low densities) and leads to
the exotic boson pairs with the d-wave symmetry. Though in most
systems, the effect of this four-site ring exchange interaction is
much smaller comparing to the two-site hopping amplitudes  because
it usually comes from the fourth order perturbation, below we would
show that we can realize the Hamiltonian as well as the
corresponding parameter region ($V\gg K \gg |t|>0$) in the BEC zone
of cold Fermi gases loaded in a two-dimensional (2D) spin-dependent
optical. Similar model without the NN repulsive interaction has been
proposed to study the exotic phases in cold atom system such as the
deconfined phase\cite{Buchler} or Bose-metal phase
\cite{Fisher,Sheng,Feiguin}. However, as we will show below, the
strong NN repulsive interaction in Eq.({\ref{EQ1}}) makes the novel
Bose-metal phase unstable.

First, we would discuss the experimental realization of our
Hamiltonian.(1) as well as the corresponding parameter regions. Most
of above discussion are based on the parameter region: $V\gg K \gg
|t|>0$ in Hamiltonian.({\ref{EQ1}}). However, in most systems, no
matter in solid physics or cold atom physics,  the effect of this
four-site ring exchange interaction is much smaller comparing to the
two-site hopping amplitudes because it usually comes from the fourth
order perturbation. Below we would show that not only the
Hamiltonian.({\ref{EQ1}}) but also the corresponding parameter
region ($V\gg K \gg |t|>0$) could be realized in the BEC zone of
cold Fermi gases loaded in a two-dimensional (2D) spin-dependent
optical\cite{Mandel}. A fermionic Hubbard model with spin-dependent
hopping has been proposed\cite{Liu}, by tuning the lasers between
hypefine structure levels of $^{40}K$ atoms. The tunneling matrix
elements for the two spin components are spin-dependent and can be
tuned with different anisotropy. In our case, we tune the hopping so
that spin up $|\uparrow\rangle$ atoms prefer to hop along the x axis
and spin down $|\downarrow\rangle$ atoms prefer to hop along the y
axis, which means there are two kind of typical hopping amplitude
$t_\sigma$ and $t_\pi$. $t_\sigma$ represents the hopping amplitude
for the $|\uparrow\rangle$ ($|\downarrow\rangle$) fermions along x
(y) axis, while $t_\pi$ denotes the hopping amplitude for the
$|\uparrow\rangle$ ($|\downarrow\rangle$) fermions along y (x) axis
(as shown in Fig.\ref{fig1}(a)). The ratio $\delta=t_\pi/t_\sigma$
can be tuned experimentally and we choose the high anisotropic
condition: $t_\pi \ll t_\sigma$. In addition, we can use Feshbach
resonances to manipulate the interactions and adjust it from
repulsive to attractive. The unconventional pairing with attractive
interaction in the anisotropic spin-dependent optical lattice has
been analyzed recently\cite{Feiguin}.

We load the fermions into the 2D spin-dependent optical lattice
defined above. Then we use Feshbach resonances to make the two
fermions occupying the same sites binding together to form a bosonic
molecule. Obviously this molecule is hard-core in nature. We assume
that the binding energy is large enough that all fermions are
tightly bound into bosonic molecule and the system enter a BEC zone.
Next we will analyze the dynamics and interactions of these new
bosons to show how can we construct the Hamiltonian ({\ref{EQ1}}) as
well as the corresponding parameter region in this spin-dependent
optical lattice.

As shown in Fig.\ref{fig1}, both the two-site hopping and
interaction involve the second order perturbation via a virtual
process. Taking the hopping term for example (Fig.1 (b)), from the
standard second order perturbation theory we can obtain the
effective hopping amplitude of the boson:
 $t=-t_\sigma t_\pi/U$,  $t_\sigma$ and
$t_\pi$ has been defined above. U is the binding energy for a
bosonic molecule formed by two fermions. Similarly, the virtual
process in Fig.1 (c) plays a role similar to the term $VS^z_iS^z_j$
($V=t_\sigma^2/U>0$) in the spin model, which means the NN repulsive
interaction $Vn_in_j$ in the boson language (those terms
proportional to $n_i$ are absorbed into the chemical potential).
Because of the strong anisotropic hopping of the fermions in the
spin-dependent optical lattice: $t_\pi\ll t_\sigma$, we have $V\gg
t$.

The virtual process  shown in Fig.1(d) is the leading term from the
fourth order perturbation. Apparently it involves four sites within
a plaquette and thus results in a ring exchange interaction in
Eq.\ref{EQ1}. From the standard perturbation theory, we can get
$K=t_\sigma^4/U^3>0$, which is much larger than the contribution of
all the other four-site virtual processes. Due to the strongly
anisotropic hopping of the fermions in the spin-dependent lattice,
it is possible to adjust the parameters
($\delta=t_\pi/t_\sigma\rightarrow 0$) in Hamiltonian (\ref{EQ1}) to
satisfy $K\gg t$ , which means that the pair binding energy is much
larger than the single-particle hopping energy.

\begin{figure}[htb]
\includegraphics[width=8.5cm,bb=2 14 209 113]
{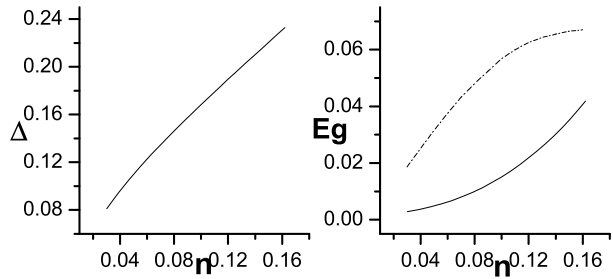} \caption{The mean-field result at zero temperature (we
set K=1 and t=0.1). The dependence of (a) energy gap $\Delta$ and
(b) the energy of the ground state  $E_g$ of the d-wave pairing
phase (solid line), and single particle BEC (dashed line) on the
total particle number.} \label{fig2}
\end{figure}

To define the symmetry of a pair, we introduce a local operator
$D_p$, which rotates the four sites within the plaquette p
cyclically by an angle $\pi/2$. The sign of K is important for the
symmetry of this bosonic pair. When $K<0$, it is s-wave ($D_p=1$);
while $K>0$ is d-wave ($D_p=-1$). This can be seen with just one
plaquette. Loading two bosons into one plaquette: due to the strong
NN repulsive interaction, the only two possible configurations in
the plaquette is two bosons occupying the diagonal sites. There are
two eigenstates of the ring-exchange term, denoted as $|s\rangle$
and $|d\rangle$(Fig.\ref{fig1}(e)), with eigenvalues K and -K,
respectively. We neglect the single particle hopping term because
$K\gg t$. Notice that when $K>0$, the ground state of this plaquette
is a d-wave state . At low densities, since the pair binding energy
is much larger than the single particle hopping energy, the bosons
prefer to move as pairs rather than hopping independently. At low
density, these bosonic pairs with d-wave symmetry will condensate to
form a d-wave superfluid.

A natural question arose here is whether the ring exchange
interaction would make the system to form bound state with more than
two bosons? Without the NN repulsive interaction, this is true: at
low density the ring exchange, just like an attractive potential,
causes not only two but many bosons to clump together, which leads
to phase separation \cite{Rousseau} between an isolated boson metal
clusters and vacuum. In our case, however, the strong repulsive
interaction will make the isolated clusters consisted of more than
two bosons unstable and break up into many bosonic pairs.

At a low density, the interaction between the pairs is not
important, thus, we can analyze the problem using the mean field
theory \cite{Bendjama}. We introduce a d-wave bosonic pair order
parameter to decouple the four-site ring exchange interaction in our
original Hamiltonian Eq. (\ref{EQ1}): $\langle a^\dag_1
a^\dag_3\rangle=-\langle a^\dag_2 a^\dag_4\rangle=\Delta$, where the
minus sign is due to the d-wave character. As we concentrate on the
low-density limit, the hard-core constraint is expected to be
irrelevant. We decouple the NN interaction by the Hartree-Fock
approximation: $Vn_in_j=V\langle n_i\rangle n_j+n_i\langle
n_j\rangle-V\langle n_i\rangle\langle n_j\rangle$. The Hamiltonian
in Eq.\ref{EQ1} can be rewritten as:

\begin{equation}
H=\sum_{\bf k} \xi_k a^\dag_k a_k+\Delta_{\bf k} a_ka_{-k}-Vn^2
+h.c+2K\Delta^2\label{EQ2}
\end{equation}
with
\begin{eqnarray}
\nonumber\Delta_k=2K\Delta\sin k_x\sin k_y,\\
\nonumber\xi_k=2t(\cos k_x+\cos k_y)-\mu+2Vn .
\end{eqnarray}
where n is the average value of the total particle number. The mean
field Hamiltonian (\ref{EQ2}) is diagonalized by using the
Bogoliubov transformation for bosons and we obtain the energy
spectrum: $E_k=\sqrt{(\xi_k/2)^2-|\Delta_k|^2}$. We focus on the
zero temperature case and the ground state energy is given as:
$E_g=\sum_{\bf k} E_k+2\Delta^2K+\mu/2-Vn-Vn^2$. The sum is over all
the $\bf k$ in the first Brillione zone. The self-consistent
equations are:
\begin{eqnarray}
n=-\frac 12+\frac 1{4L}\sum_{\bf k}\frac{\xi_{\bf k}}{E_{\bf k}},\\
1=\frac {K}L\sum_{\bf k}\frac{\sin^2k_x\sin^2k_y}{E_{\bf k}}.
\end{eqnarray}

\begin{figure}[htb]
\includegraphics[width=7.1cm,bb=1 450 367 639]
{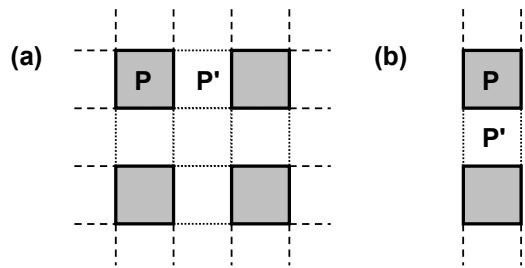} \caption{The optical superlattice for (a) square
lattice and (b) ladder with periodic boundary condition }
\label{fig3}
\end{figure}

Since we are dealing with a bosonic system, it is possible that
another BEC state with single particle condensation will compete
with our d-wave pairing state. In this case, Eq.(3) should be
replaced by $n=n_c-\frac{\partial E_g}{\partial \mu}$, where $n_c$
is the density of bosons with single particle condensation. We find
that at least in our parameter regime $t\ll K\ll V$, there is no
positive self-consistent solution for $n_c$.  The absence of single
particle condensation has been observed previously in a bosonic
system with correlated hopping\cite{Bendjama}. It is shown that the
bosons prefer to pairing with each other due to the strong effective
attractive interaction. To clarify this point, we also calculate the
energy of the single particle BEC state without d-wave pairing via
the standard Bogoliubov approximation: $a_i=\sqrt{n_c}+\delta a_i$
to decouple the Hamiltonian.(1). The result is shown in Fig.2 (we
set K=1 and t=0.1). Notice that at least at low densities with $K\gg
t$, the ground state energy of d-wave pairing state is always lower
than that of the single particle BEC state.

\begin{figure}[htb]
\includegraphics[width=8.5cm,bb=5 23 301 175]
{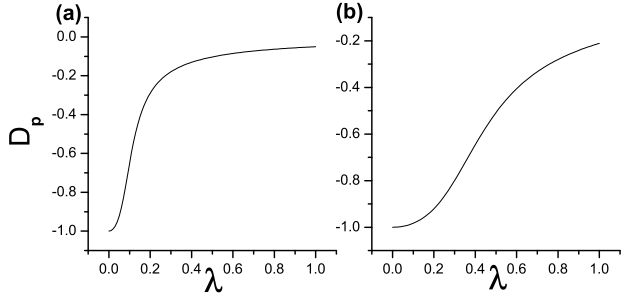} \caption{The dependence of $\langle D_p\rangle$ on
$\lambda$ for a (a) $4\times4$ square superlattice (b) $2\times 8$
ladder superlattice with periodic boundary condition} \label{fig4}
\end{figure}

Next£¬we would turn to another limit, when the filling factor is
$1/2$. In this case, the strong NN repulsive interaction induces a
conventional $(\pi,\pi)$ density wave phase, rather than the boson
metal or d-wave bosonic pairs. However, if we load the half filling
bosons into a 2D optical superlattice, it is possible to recover the
d-wave symmetry within plaquette and lead to a d-wave Mott insulator
\cite{Yao,Trebst,Peterson}. Next we will clarify this point by the
exact diagonalization (ED) of the small size systems. We classify
all the plaquettes in the superlattice as two classes: P (grey
plaquette in Fig.\ref{fig3})  and P' (white plaquette) and the
Hamiltonian in this case is given by:
\begin{equation}
H_s=\sum_{\Box\in P} H_0+\lambda \sum_{\Box\in P'} H_0,
\end{equation}
where $H_0$ is the Hamiltonian defined by Eq.(\ref{EQ1}) in one
plaquette and $0<\lambda<1$. Notice that when $\lambda\ll 1$, the
situation is similar to that in a single plaquette, the system forms
a d-wave Mott insulator with $\langle D_p\rangle\approx-1$. When
$\lambda\approx 1$, the ground state should be a $(\pi,\pi)$ DW with
$\langle D_p\rangle=0$ due to the strong repulsive NN interaction.
$\langle D_p\rangle$ is the expectation value of the rotating
operator defined above to measure the d-wave symmetry within one
plaquette. We calculate $\langle D_p\rangle$ in a $4\times4$
superlattice (Fig.\ref{fig4}(a)) and a  $2\times8$ ladder
(Fig.\ref{fig4}(b)) with periodic boundary conditions, to show how
it changes when we increase $\lambda$ from 0 to 1.

\begin{figure}[htb]
\includegraphics[width=8.5cm, bb=26 310 532 699]
{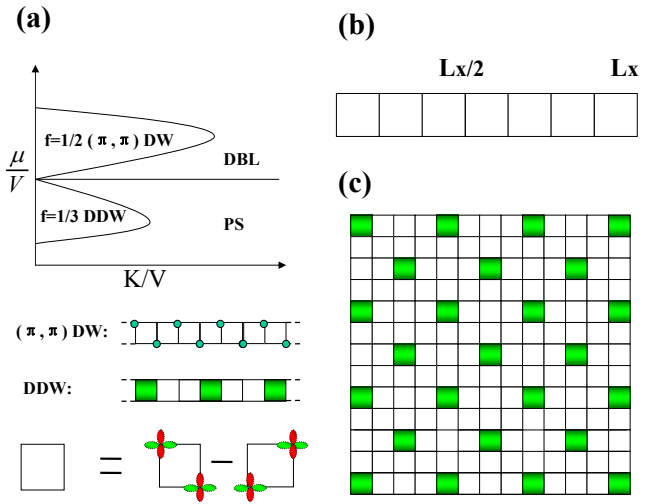} \caption{(a) Sketch phase diagram of the model in
Hamiltonian (1) in a two-leg ladder ($t\approx 0$) (b)The structure
of the DDW state in a  two-leg ladder (1/3 filling) . (c) The
structure of the DDW state in a 2D system (quarter-filling)}
\label{fig5}
\end{figure}

At some particular filling factor (f=1/3 for the two-leg ladder
system and 1/4 for the 2D system), a novel phase emerges. The ring
exchange interaction makes two boson prefer to form a d-wave pair,
while the strong NN interaction prevents two pairs from being too
close. The conspiracy of them makes these d-wave pairs localized and
localized and separated as far as possible to avoid the strong NN
interaction. The crystallization of these d-wave pairs leads to a
novel phase: d-wave density wave state, which preserves the d-wave
symmetry within plaquette and spontaneously breaks the translational
symmetry, as shown in Fig.\ref{fig5}(b) (two-leg ladder) and
Fig.\ref{fig5}(c) (2D). Notice that unlike the half filling case,
the translational symmetry breaking is spontaneously in DDW state
thus we don't need any superlattice structure.

Before discussing the TEBD result, we first briefly discuss the
global phase diagram of our ladder system in the limit of
$t\rightarrow 0$ in Hamiltonian.(\ref{EQ1}). First we analyze the
half filling case, in the limit $V\rightarrow 0$, it is known that
this K-only model can be mapped to a hard-core boson model and could
be solved exactly\cite{Sheng}. Its ground state is a gapless highly
correlated state of boson: d-wave Bose liquid (DBL). When $V\gg K$,
a gapped $(\pi,\pi)$ density wave (DW) state would dominate. A
quantum phase transition would occur when we increase the NN
interaction V. When the filling factor f=1/3, in the limit
$V\rightarrow 0$, we anticipate that the phase of ground state is
separated into an empty region and a half-filled region of the DBL
state. When we increase V, as analyzed above, the competition
between the NN interaction and ring exchange interaction would lead
to a DDW state, which preserves the local d-wave symmetry and breaks
the translational symmetry spontaneously. The sketch global phase
diagram is shown in Fig.\ref{fig5}(a).

Below we would study the DDW state and the properties of the quantum
phase transition in the two-leg ladder system by the TEBD method. We
focus on the case the filling factor $f\approx 1/3$. The open
boundary condition is used to artificially shift the ground state
degeneracy of DDW state due to the spontaneous translational
symmetry breaking, therefore the filling factor is not exactly 1/3
in the ladder with finite length. For example in a $2\times L_x$
ladder, the number of boson $N_b=(2L_x+2)/3$. In the thermodynamic
limit, $L_x\rightarrow \infty $, the filling factor is exactly 1/3.
Because the DBL state and the phase separation as well as  the
quantum phase transition between them have been explicitly discussed
in Ref.\cite{Sheng}, thus we would not discuss them here and mainly
focus on the DDW state and the properties of the corresponding
quantum phase transition.

\begin{figure}[htb]
\includegraphics[width=7.0cm,bb=14 14 315 384]
{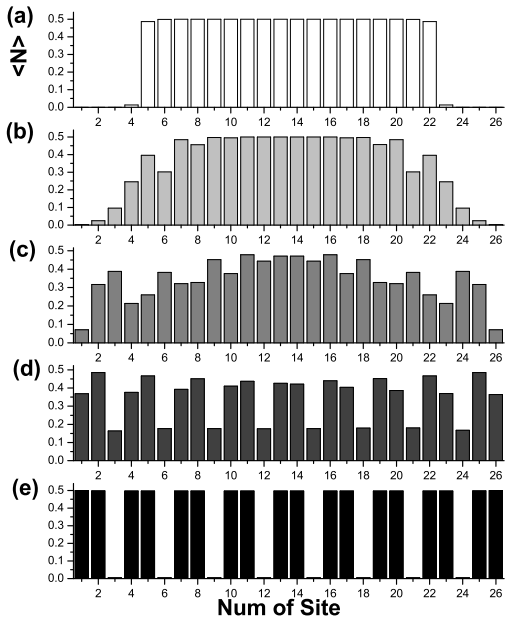} \caption{The particle distribution in the real space
in a $2\times 26$ ladder, we set t=0.1, K=1.0 and (a)V=0.01
(b)V=0.41, (c)V=0.44, (d)V=0.46, (e)V=3.0} \label{fig6}
\end{figure}

Because of the different structure of the phase separation and the
DDW state,the phase transition between them can been seen most
directly from the particle number distribution in the real place, as
shown in Fig.\ref{fig6}, where we set $t=0.1, K=1.0, L_x=26$. In the
phase separation region (V=0.05), due to the open boundary
condition, the boson prefer to get together in the center of the
ladder to form a half-filling DBL and the rest part is
empty\cite{Sheng}. Deep in the DDW state (V=3.0), there is a
three-period crystal structure of the particle number distribution.
There is a quantum phase transition between these two phases.

\begin{figure}[htb]
\includegraphics[width=8.0cm,bb=14 14 281 274]
{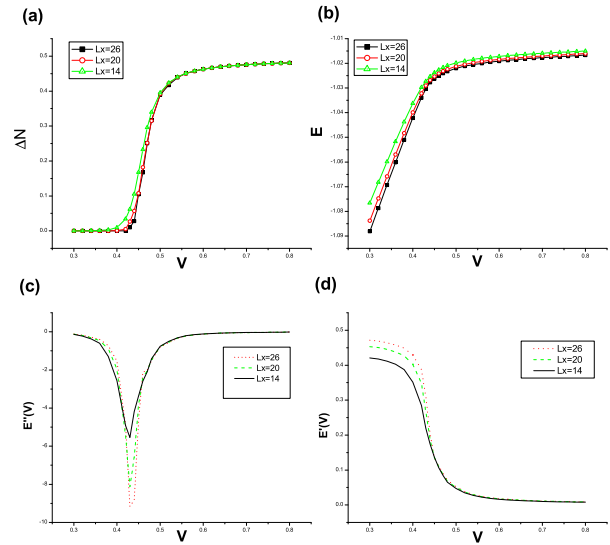} \caption{(a)The dependence of $\Delta N$ on V in a
two-leg ladder, we set t=0.1, K=1.0; (b)The average energy per boson
E $.vs.$ V; (c) $\frac{\partial^2E}{\partial V^2}$ $.vs.$ V;
(d)$\frac{\partial E}{\partial V}$ $.vs.$ V.} \label{fig7}
\end{figure}

To study the properties of the quantum phase transition, we
introduce $\Delta N=\rho(L_x/2)-\rho(L_x/2-1)$ as the order
parameter to characterize the density wave state, where $\rho(i)$
has been defined above. The dependence of $\Delta N$ on the NN
interaction V is shown in Fig.\ref{fig7}(a), where we set $t=0.1$
and $K=1.0$ for simplicity. In a perfect DDW state
($V\rightarrow\infty$), $\Delta N=0.5$, while in the opposite limit
(V=0), a phase separation means $\Delta N=0$, (in our case K/t=10,
thus the ground state in K=0 is phase separation rather than
DBL\cite{Sheng}). Fig.\ref{fig7}(a) indicate a strong signature of a
continuous quantum phase transition, instead of a simple first-order
phase transition caused by energy level crossing. To verify this
point, we also calculate the dependence of the average ground state
energy per boson (Fig. \ref{fig7}(b)), its first  as well as the
second order derivative (Fig.\ref{fig7}(c) and (d)) on the NN
interaction V. We notice that there is no discontinuity in the first
order derivative ($\frac{\partial E}{\partial V}$) while  a sharp
peak appears in the second order derivative of the ground state
energy ($\frac{\partial^2E}{\partial V^2}$), which indicates that a
second order phase transition occurs at the point $V_c=0.43\sim
0.44$.

The pairing between two bosons have recently attracted considerable
attentions, while most of previous pairing mechanisms are based on
the direct attractive interspecies interactions tuned by Feshbach
Resonance \cite{Romans, Bhaseen}, or on a three-body onsite hardcore
constraint \cite{Diehl}. All of these mechanisms are due to the
uniqueness of the cold atomic system and have no counterpart in
traditional condensed matter physics. In this paper, we proposed a
novel pairing mechanism for bosons via strong four-site ring
exchange interaction, which is also thanks to the unique feature of
the ultracold atoms in optical lattice.  Recently a proposal has
been provided to experimentally detect these bosonic
pair\cite{Menotti}, which would be helpful to detect the bosonic
pair in our case.

In summary, we propose a strongly correlated bosonic Hamiltonian
with four-site ring exchange interaction. We focus on the parameter
region $V\gg K \gg |t|>0$ and investigate the exotic phases with
d-wave symmetry emerging at different filling factors. A physical
realization of the Hamiltonian as well as the parameter region has
also been discussed.

The authors are grateful to Congjun Wu  for helpful discussion. ZC
is supported by AROW911NF0810291.  The work is also supported in
part by NSF-China, MOST-China and US-DOE-FG02-04ER46124 (XCX).


\begin{references}




\bibitem{Greiner} M. Greiner, O. Mandel, T. Esslinger, T. W. Hansch and I. Bloch. Nature. {\bf 415}, 39
(2002).

\bibitem{Jordens}R. J\"{o}rdens, N. Strohmaier, K. G\"{u}nter, H. Moritz, T.
Esslinger, Nature. {\bf 445}, 204 (2008).

\bibitem{Trebst} S. Trebst, U. Schollw\"{o}ck, M. Troyer, and P.
Zoller, Phys. Rev. Lett. {\bf 96}, 250402 (2006).





\bibitem{Peterson} M.R. Peterson, C.W. Zhang, S.T. Tewari, and S.
Das Sarma, Phys. Rev. Lett. {\bf 101}, 150406 (2008).

\bibitem{Paredes}B. Paredes and I. Bloch, Phys. Rev. A {\bf 77},
023603 (2008).

\bibitem{Hofstetter} W. Hofstetter, J.I. Cirac, P. Zoller, E. Demler, and M. D.
Lukin, Phys. Rev. Lett. {\bf 89}, 220407 (2002).

\bibitem{Honerkamp} C. Honerkamp and W. Hofstetter, Phys. Rev. Lett. {\bf 92}, 170403
(2004).

\bibitem{Buchler} H.P. B\"{u}chler, M. Hermele, S.D. Huber,
Matthew.P.A. Fisher, and P. Zoller, Phys. Rev. Lett. {\bf 95},
040402 (2005).

\bibitem{Tewari} S. Tewari, V.W. Scarola, T. Senthil and S. Das Sarma, Phys. Rev.
Lett. {\bf 97}, 200401 (2006).


\bibitem{Han} Y.J. Han, R. Raussendorf and L.M. Duan, Phys. Rev.
Lett. {\bf 98}, 150404 (2007).



\bibitem{Wu2} C.J. Wu, Phys. Rev. Lett. \textbf{100}, 200406 (2008); Phys. Rev. Lett. \textbf{101}, 186807 (2008);
C.J. Wu and S. Das Sarma, Phys. Rev. B \textbf{77}, 235107 (2008);
W.C. Lee and C.J. Wu, arXiv: 0905.1146; H. H. Hung, W.C. Lee and
C.J. Wu, arXiv: 0910.0507;

\bibitem{Zhao} E. H. Zhao and W.V. Liu, Phys. Rev. Lett. \textbf{100},
    160403 (2008).

\bibitem{Wang} L. Wang, X. Dai, S. Chen, and X.C. Xie, Phys. Rev.
    A \textbf{78}, 023603 (2008).



\bibitem{Paredes2}B. Paredes and I. Bloch, Phys. Rev. A {\bf 71},
063608 (2005).

\bibitem{Damski}L. Santos, M. A. Baranov, J. I. Cirac, H.-U. Everts, H. Fehrmann, and M.
Lewenstein, Phys. Rev. Lett. {\bf 95}, 060403 (2005); Phys. Rev. A
{\bf 72}, 053612 (2005).

\bibitem{Goodman} T. Goodman and L.M. Duan, Phys. Rev. A {\bf 74}, 052711
(2006).

\bibitem{Yao}H. Yao, W.F. Tsai and S.A. Kivelson, Phys. Rev. B {\bf
76}, 161104(R) (2007).

\bibitem{Rey} A.M. Rey, R. Sensarma, S. Foelling, M. Greiner, E.
Demler, and M.D. Lukin, arXiv:0806.0166




\bibitem{Scalapino} D.J. Scalapino, S.A. Trugman, Philos. Mag. B
{\bf 74}, 607 (1996);  K. Le Hur and T. M. Rice,  Annals of Physics
{\bf 324}, 1452 (2009)



\bibitem{Fisher} A. Paramekanti, L. Balents, and M.P.A. Fisher,
Phys. Rev. B {\bf 66}, 054526 (2002); O.I. Motrunich and M.P.A.
Fisher, Phys. Rev. B {\bf 75}, 235116 (2007);

\bibitem{Sheng} D.N. Sheng, O.I. Motrunich, S. Trebst, E. Gull, and M.P.A.
Fisher, Phys. Rev. B {\bf 78}, 054520 (2008)

\bibitem{Feiguin}A. E. Feiguin and M. P. A. Fisher, Phys. Rev. Lett. {\bf 103}, 025303
(2009) .


\bibitem{Mandel}O. Mandel et al., Nature (London) 425, 937 (2003); Phys. Rev.
Lett. 91, 010407 (2003).

\bibitem{Liu} W. Vincent Liu, F. Wilczek and P. Zoller,  Phys. Rev. A {\bf 70}, 033603 (2004).


\bibitem{Rousseau} V.G. Rousseau, R.T. Scalettar, and G.G. Batrouni,
Phys. Rev. B {\bf 72} 054524 (2005); V.G. Rousseau,  G.G. Batrouni
and R.T. Scalettar, Phys. Rev. Lett. {\bf 93} 110404 (2004);

\bibitem{Bendjama} R. Bendjama, B. Kumar, and F. Mila, Phys. Rev.
Lett. {\bf 95}, 110406 (2005);

\bibitem{Vidal}G. Vidal, Phys. Rev. Lett. {\bf 91}, 147902 (2003);
Phys. Rev. Lett. {\bf 93}, 040502 (2004).


\bibitem{Romans} M.W. J. Romans, R. A. Duine, Subir Sachdev, and H.T.C.
Stoof, Phys. Rev. Lett. {\bf 93}, 020405 (2004).

\bibitem{Bhaseen} M. J. Bhaseen, A. O. Silver, M. Hohenadler, B. D.
Simons, Phys. Rev. Lett. {\bf 103}, 265302 (2009).

\bibitem{Diehl} S. Diehl, M. Baranov, A. J. Daley, P. Zoller, arXiv:
0910.1859 (2009)

\bibitem{Menotti}C. Menotti and S. Stringari, arXiv: 0912.4452
(2009)


























\end{references}
\end{document}